\documentclass{PoS}
\usepackage[T1]{fontenc} % if needed
\usepackage[utf8]{inputenc}
\usepackage{subcaption}
\usepackage{caption}
\usepackage{graphicx}
\usepackage{color}
\usepackage{wrapfig}
\usepackage{amsmath}
\usepackage{mathtools}
\title{LHC constraints on the minimal Dirac gaugino model.}

\ShortTitle{LHC constraints on the MDGSSM.}

\author{Guillaume Chalons$^{a}$, Mark Goodsell$^{b}$, Sabine Kraml$^{a}$, \speaker{Humberto Reyes-Gonz\'{a}lez}$^{a}$, Sophie L. Williamson$^{b,c}$.\\
\llap{$^a$}Laboratoire de Physique Subatomique et de Cosmologie, Universit\'{e} Grenoble-Alpes, \\ CNRS/IN2P3, 53 Avenue des Martyrs, F-38026 Grenoble, France.\\
\llap{$^b$}Laboratoire de Physique Th\'{e}orique et Hautes Energies (LPTHE), \\ UMR 7589, Sorbonne Universit\'{e} et CNRS, 4 place Jussieu, 75252 Paris Cedex 05, France.\\
\llap{$^c$} Institute for Theoretical Physics, Karlsruhe Institute of Technology, \\
Wolfgang-Gaede-Str. 1, 76131 Karlsruhe, Germany.

E-mail: \email{chalons@lpsc.in2p3.fr}, \email{goodsell@lpthe.jussieu.fr}, \email{sabine.kraml@lpsc.in2p3.fr}, \email{gonzalez@lpsc.in2p3.fr}, \email{sophie.williamson@kit.edu}}   

\abstract{Most SUSY searches at the LHC are optimised for the MSSM, where gauginos are Majorana particles. By introducing Dirac gauginos, we obtain an enriched phenomenology, from which considerable differences in the LHC signatures and limits are expected as compared to the MSSM. Concretely, in the minimal Dirac gaugino model (MDGSSM) we have six neutralino and three chargino states. Moreover, production cross sections are enhanced for gluinos, while for squarks they are suppressed. In this contribution, we explore the consequences of the current LHC limits on gluinos and squarks in this model.}

\FullConference{
European Physical Society Conference on High Energy Physics - EPS-HEP2019 -\\
			10-17 July, 2019\\
			Ghent, Belgium}

\begin{document}

\section{Introduction.}

Originally proposed to allow the gluino to be massive~\cite{Fayet:1978qc}, models with Dirac Gauginos (DG) have interesting features such as providing a tree level boost to the Higgs mass, being associable to $N=2$ supersymmetry, the posibility of preserving R-symmetry and increasing naturalness. See \cite{Benakli:2011vb}, for a concise review on Dirac Gauginos.

\begin{table}[htb]
\begin{center}
\small
\begin{tabular}{|c|c|c|c|c|c|}
\hline
Names  &                 & Spin 0                  & Spin 1/2 & Spin 1 & $SU(3)$, $SU(2)$, $U(1)_Y$ \\ 
\hline

Quarks  & $\mathbf{Q}$   & $\tilde{Q}=(\tilde{u}_L,\tilde{d}_L)$  & $(u_L,d_L)$ & & (\textbf{3}, \textbf{2}, 1/6) \\ 
   & $\mathbf{u^c}$ & $\tilde{u}^c_L$              & $u^c_L$     & & ($\overline{\textbf{3}}$, \textbf{1}, -2/3) \\ 
($\times 3$ families) & $\mathbf{d^c}$ & $\tilde{d}^c_L$     & $u^c_L$     & & ($\overline{\textbf{3}}$, \textbf{1}, 1/3)  \\ 
\hline
Leptons & $\mathbf{L}$ & ($\tilde{\nu}_{eL}$,$\tilde{e}_L$) & $(\nu_{eL},e_L)$ & & (\textbf{1}, \textbf{2}, -1/2) \\ 
($\times 3$ families) & $\mathbf{e^c}$ & $\tilde{e}^c_L$    & $e^c_L$          & & (\textbf{1}, \textbf{1}, 1)  \\ 
\hline
Higgs & $\mathbf{H_u}$ & $(H_u^+ , H_u^0)$ & $(\tilde{H}_u^+ , \tilde{H}_u^0)$ & & (\textbf{1}, \textbf{2}, 1/2)  \\ 
  & $\mathbf{H_d}$ & $(H_d^0 , H_d^-)$ & $(\tilde{H}_d^0 , \tilde{H}_d^-)$ & & (\textbf{1}, \textbf{2}, -1/2) \\
\hline
Gluons & $\mathbf{W_{3\alpha}}$ & & $\tilde{g}_{\alpha}$          & $g$              & (\textbf{8}, \textbf{1}, 0) \\ 

W    & $\mathbf{W_{2\alpha}}$ & &  $ \tilde{W}^{\pm} , \tilde{W}^{0}$ & $W^{\pm} , W^0$  & (\textbf{1}, \textbf{3}, 0) \\ 

B    & $\mathbf{W_{1\alpha}}$ & &  $ \tilde{B}$                    & $B$              & (\textbf{1}, \textbf{1}, 0 ) \\ 
\hline
\hline
\textcolor{red}{DG-octet} &\textcolor{red}{ $\mathbf{O_g}$} &  \textcolor{red}{$O_g $}  &   \textcolor{red}{$  \tilde{g}'$ } &  & \textcolor{red}{(\textbf{8}, \textbf{1}, 0)} \\ 

\textcolor{red}{DG-triplet} & \textcolor{red}{$\mathbf{T}$} & \textcolor{red}{$\{T^0, T^{\pm}\}$} & \textcolor{red}{ $\{\tilde{W}'^{\pm},\tilde{W}'^{0}\}$}&  & \textcolor{red}{(\textbf{1},\textbf{3}, 0 )}\\  

\textcolor{red}{DG-singlet}  &\textcolor{red}{ $\mathbf{S}$}& \textcolor{red}{$S$} & \textcolor{red}{$  \tilde{B}'$ }  &  & \textcolor{red}{(\textbf{1}, \textbf{1}, 0 ) }\\ 
\hline
\end{tabular}
\caption{Chiral and gauge multiplet fields in the model. The red coloured section corresponds to the new chiral multiplets that completes the DG model. The rest are the usual multiplets found in the MSSM.}
\label{tab:content}
\end{center}
\end{table}

To introduce Dirac masses for the gauginos, we need to add a Weyl fermion in the adjoint representation of each gauge group; these are embedded in chiral superfields $\mathbf{S}, \mathbf{T}, \mathbf{O}$ which are respectively a singlet, triplet and octet,  and carry zero R-charge. The resulting field content is summarised in Table~\ref{tab:content}. The singlet and triplet fields can have new superpotential couplings with the Higgs, 
\begin{align}
W\supset \lambda_S \bold{S} \, \bold{H_u} \cdot \bold{H_d} + 2 \lambda_T \, \bold{H_d} \cdot \bold{T} \bold{H_u} \label{EQ:WNeq2}\,. 
\end{align}

\begin{figure}%{r}{.57\textwidth}
\centering
\includegraphics[width=0.57\textwidth]{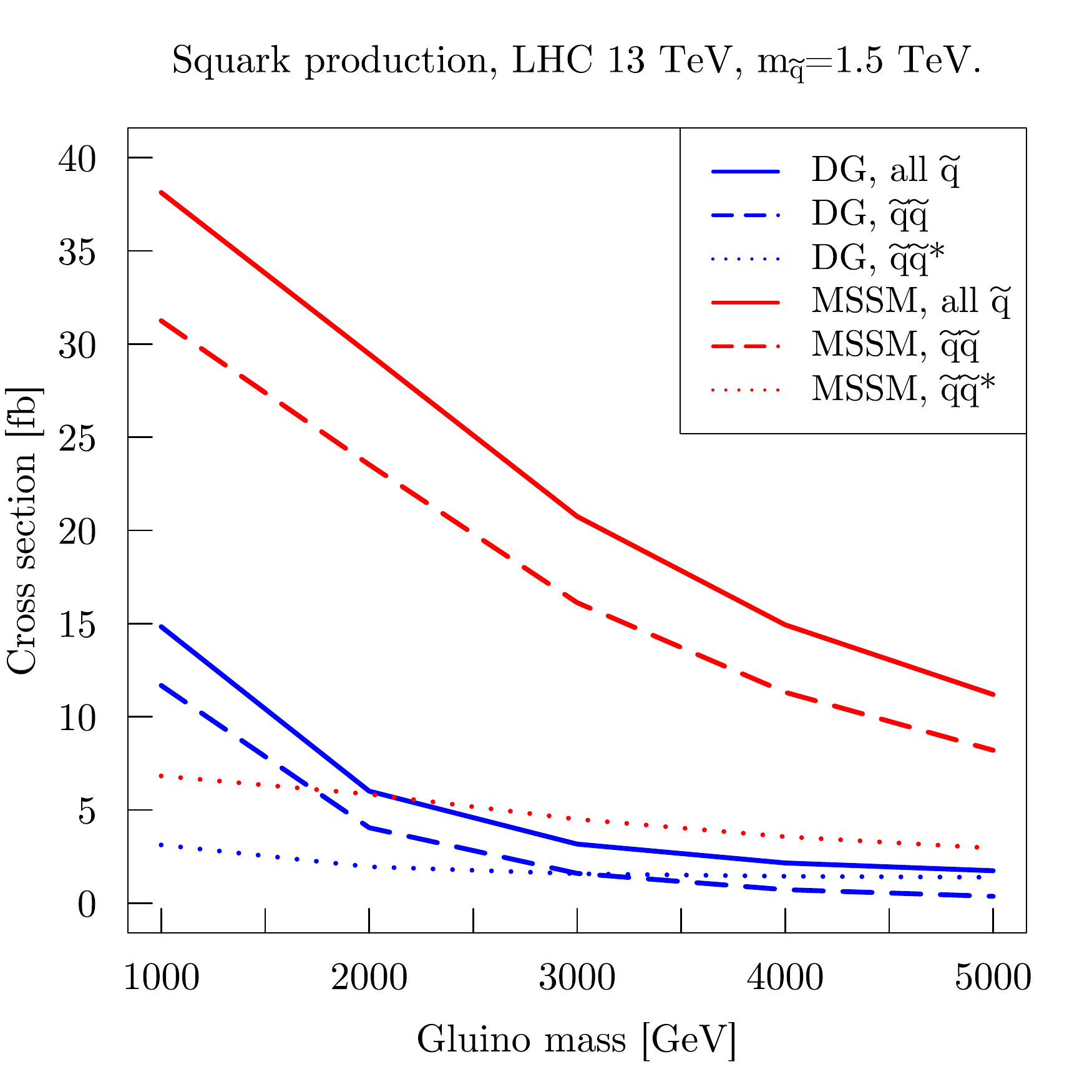}
\caption{\label{fig:squark-production}
Squark production cross-sections at leading order (LO) for the 13 TeV LHC as a function of the gluino mass in the MSSM (in red) and in 
the DG case (in blue).}
\label{fig:cross_sections}
\end{figure}
In DG models, different limits from collider searches can be expected when looking at the production cross sections of squarks and gluinos. For squarks, pair production is suppressed due to the absence of a chirality flip (see Fig.~\ref{fig:cross_sections}), whereas gluino-pair production is enhanced because of an enlarged number of degrees of freedom compared to the MSSM. We can also expect different limits from the fact that DG models have a more complex electroweak-ino spectrum, with six neutralino and three chargino mass eigenstates (as compared to four and two, respectively, in the MSSM), which may appear in gluino and squark cascade decays.

In the following, we show limits on the masses of the gluinos and squarks in the MDGSSM obtained from reinterpreting LHC results, using the full recasting approach. A complete description of this work can be found in \cite{Chalons:2018gez}.

\section{Deriving limits on the masses of gluinos and squarks. }

In this contribution, we will focus on four benchmark scenarios, denoted as DG1, DG2, DG3 and DG4. In all of them, $m_{DY}=200$ GeV, $\mu=400$ GeV, $\tan\beta=2$ and $\lambda_T$=2, where $m_{DY}$ is the soft mass term of the Dirac binos, $\mu$ is the higgsino mass term and $\tan\beta\equiv v_u/v_d$. Regarding the soft mass term of the Dirac winos, in the DG1, DG2 and DG3 scenarios, it is fixed to $m_{D2}=500$ GeV, so as to have light winos, while in DG4 $m_{D2}=1175$ GeV, thus obtaining heavy winos. This translates, for all cases, into a hierarchical spectrum of bino ($\tilde\chi^0_{1,2}$)-, higgsino ($\tilde\chi^0_{3,4}$)- and wino ($\tilde\chi^0_{5,6}$)-like states with masses of about 200, 400 and 500~GeV (for DG1, DG2 and DG3) or 1226~GeV (for DG4), respectively. 

\begin{figure}\centering
\includegraphics[width=0.5\textwidth]{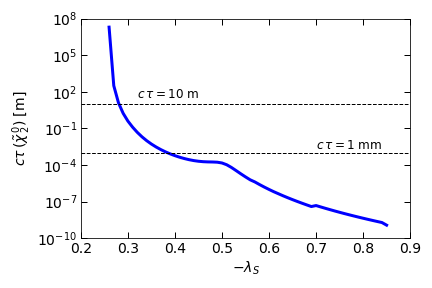}%
\includegraphics[width=0.5\textwidth]{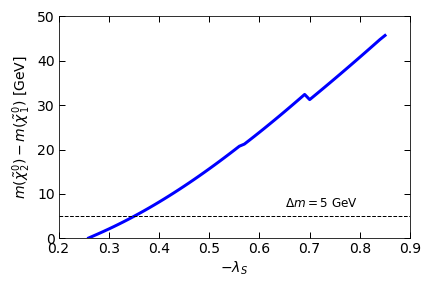}
\caption{Influence of $\lambda_S$ on the mass splitting between the two bino-like mass eigenstates $\tilde\chi^0_{1,2}$ (left) and 
on the lifetime of the $\tilde\chi^0_{2}$ (right) .}
\label{fig:benchmarks}
\end{figure}

With this configuration, the two lightest neutralinos, $\tilde\chi^0_{1,2}$, will be mostly bino states with a mass splitting given by $m_{\tilde\chi^0_{2}}-m_{\tilde\chi^0_{1}}=|\frac{M_Z^2\sin^2\theta_W}{\mu}\frac{2\lambda_S^2-g_Y^2}{g^2_Y}\sin 2\beta|$. The influence of $\lambda_S$ on the mass splitting of the bino like states and, subsequently, on the mean decay length of $\tilde\chi^0_{2}$, is shown in Fig.~\ref{fig:benchmarks}; this was important for choosing our benchmark scenarios. When $|\lambda_S|$ is small, the mass difference is in the MeV range, and the mean decay length of $\tilde\chi^0_{2}$ can be of the order of kilometers, so it will appear as a co-LSP. With this in mind, we have chosen $\lambda_S=-0.27$, for the DG1 scenario. When  $|\lambda_S|$ is large, the mean decay length of $\tilde\chi^0_{2}$ is so small that it decays promptly; this is the case in DG2 and DG3, where $\lambda_S=-0.74$, and in DG4 where $\lambda_S=-0.79$.

Finally, the masses of gluinos and squarks were treated as free parameters, while the masses of the 3rd generation squarks were adjusted such that $m_{h_1}\in [123,\,127]$~GeV. The rest of the particle content is decoupled.

To derive constraints on the gluinos and squarks of the model, we implemented (see \cite{ma5:recast}) the ‘Meff-based’ signal regions of the  ATLAS multijet search~\cite{Aaboud:2017vwy}  for squarks and gluinos in final states with 2-6 jets and large missing transverse momentum, using \mbox{36~fb$^{-1}$} of $\sqrt{s} = 13$~TeV $pp$ collision data, in {\tt MadAnalysis\,5}~\cite{ Dumont:2014tja}. 

We scanned over gluino and light-flavor squark masses for the DG1, DG2, DG3 and DG4 scenarios. For each scan point, we simulated 30K events with {\tt MadGraph5\_aMC@NLO}~\cite{Alwall:2014hca}, with parton showering and hadronization done in {\tt Pythia\,8.2}~\cite{Sjostrand:2014zea}
and the simulation of the ATLAS detector with  {\tt Delphes\,3}~\cite{deFavereau:2013fsa}. Afterwards, the events were analysed with {\tt MadAnalysis\,5} and  exclusion confidence levels (CL)  were computed. Finally, for limit setting, since signal regions are inclusive, only the ``best'' (i.e.\ the statistically most sensitive) were used.

\begin{figure}%{r}{0.6\textwidth}
\centering
\includegraphics[width=.7\textwidth]{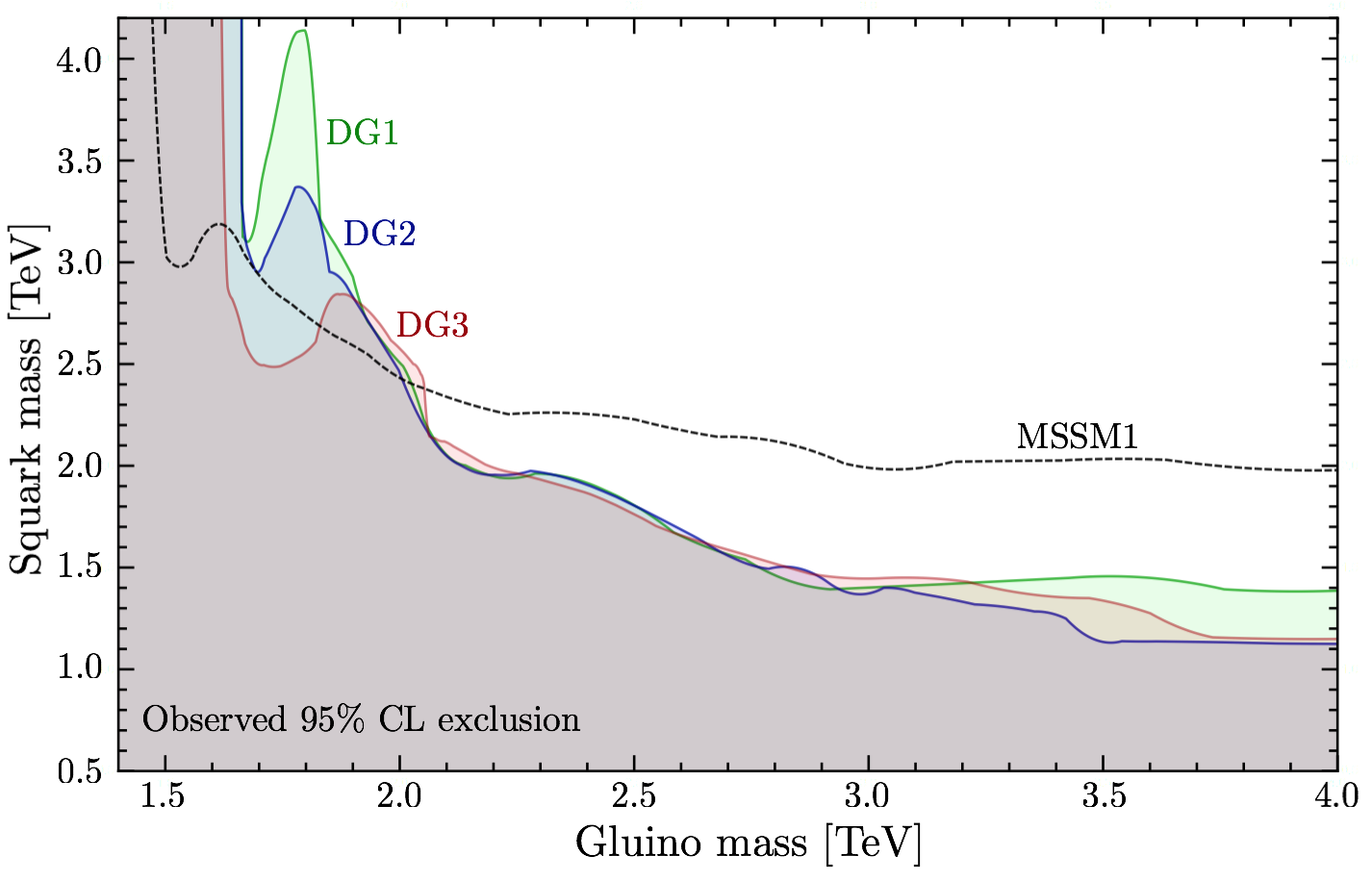}
\caption{95\% CL exclusion limits in the gluino vs.\ squark mass plane for DG1 (green), DG2 (blue) and DG3 (red) contrasted with MSSM1 (black dashed line), derived from the recasting of ~\cite{Aaboud:2017vwy}.}\label{fig:1a}		
\end{figure}

Figure~\ref{fig:1a} shows the 95\% CL exclusion lines in the gluino vs. squark mass plane for the  benchmark scenarios with light winos,  and for MSSM1, an MSSM scenario equivalent to DG1. In the region $m_{\tilde q}>m_{\tilde g}$ we found a robust limit of $m_{\tilde g}\gtrsim 1.65$~TeV when squarks are very heavy, in all cases. In the region  $m_{\tilde g}>m_{\tilde q}$, for DG2 and DG3 we obtained a squark limit of $m_{\tilde q}\gtrsim 1.1$~TeV, while for DG1 the limit reaches $m_{\tilde q}\gtrsim 1.4$~TeV. The difference is a consequence of the $\tilde\chi^0_2\to Z^*\tilde\chi^0_1$ decays, which are present in DG3 and DG2 but not in DG1.

We also observe different ``dips'' in the exclusion contours for the different benchmark scenarios which originate from a switch of the best signal region from  6j-Meff-1800 (6~jets, $M_{\rm eff}>1800$~GeV) to  6j-Meff-2600 (6~jets, $M_{\rm eff}>2600$~GeV) at different, but close, values of gluino mass for each scenario.
Futhermore, looking at the exclusion line for MSSM1,  we see the expected $\sim 200$~GeV lower gluino mass limit for an MSSM scenario and a stronger squark mass limit when gluinos are heavy, still reaching $m_{\tilde q}\gtrsim 2$~TeV for 4~TeV gluinos, as Majorana gluinos decouple very slowly.  
\begin{figure}%{r}{0.6\textwidth}
\centering
\includegraphics[width=.7\textwidth]{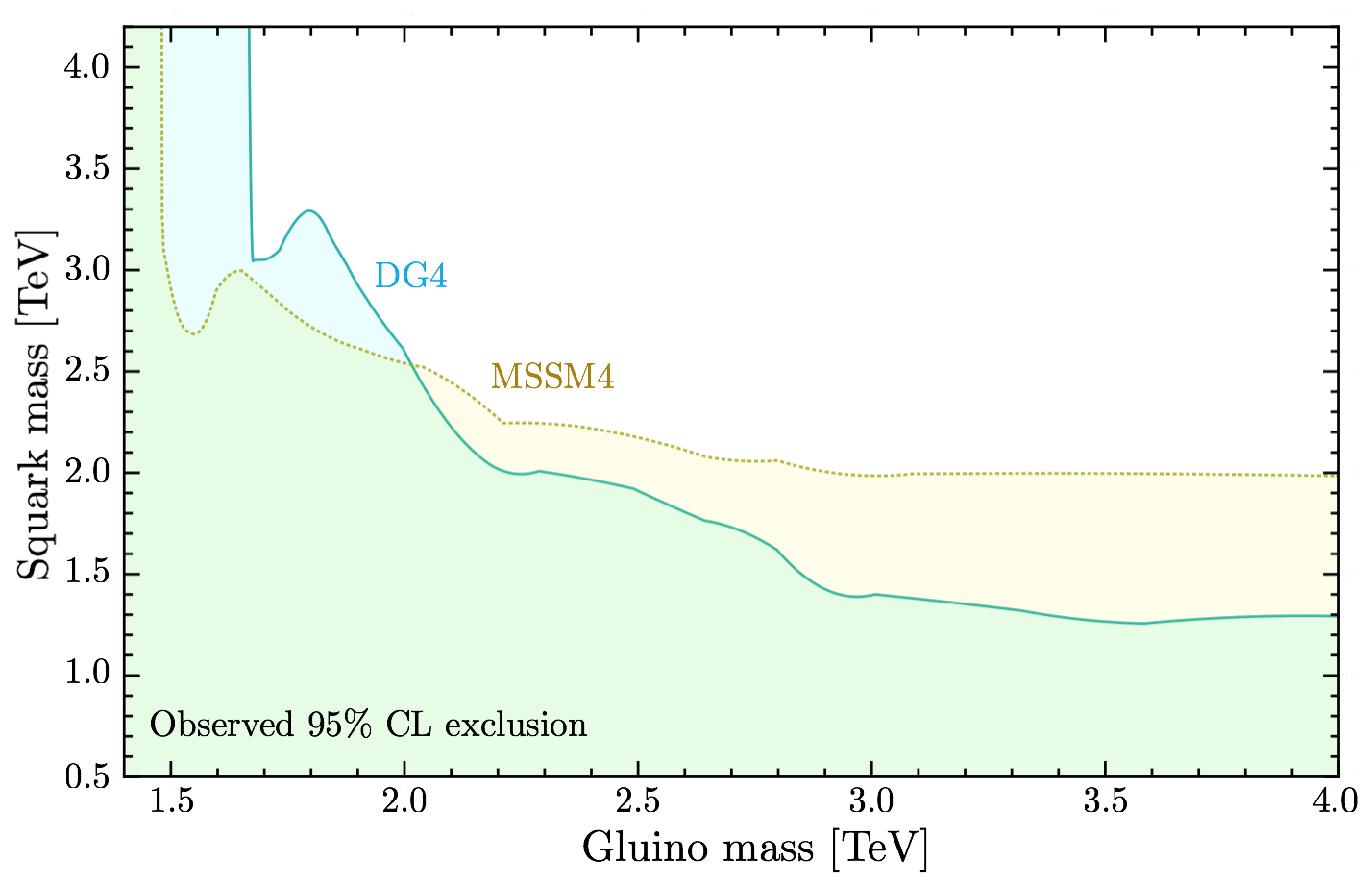}
\caption{95\% CL exclusion limits in the gluino vs.\ squark mass plane for DG4 (in blue) compared with MSSM4 (in yellow), derived from the recasting of ~\cite{Aaboud:2017vwy}.}\label{fig:2a}		
\end{figure}

Finally, turning to the heavy winos scenario, Fig.~\ref{fig:2a} show the 95\% CL exclusion limits in the gluino vs. squark mass plane for the DG4 and its MSSM equivalent MSSM4 scenarios. Interestingly, when comparing these limits with the ones for DG2 and DG3, we find very similar results; the main difference is a small increase of the squark limit by about 100--200 GeV, when winos are heavy. Regarding DG4 compared to MSSM4, the same arguments as for the light wino cases apply.

\section{Conclusions.}
As expected, by allowing Dirac gauginos we obtained LHC limits  that differ from the ones of the MSSM. The differences found come mainly from suppressed squark production, enhanced gluino production and a more complicated electroweakino spectrum. Also interesting is the possibility of a long lived $\tilde\chi^0_{2}$; a follow-up project aims to study this kind of scenario, in the light of Long Lived Particle searches. 

In general, we think that studying the MDGSSM and other non-minimal SUSY models is a well motivated endeavor, and we hope that this contribution provides a good example of its usefulness.

\section*{Acknowledgements.}
This work was supported in part by the IN2P3 project ``Th\'{e}orie --LHCiTools''. It has also been done within the Labex ILP (reference
ANR-10-LABX-63) part of the Idex SUPER, and received financial state aid managed by the Agence Nationale de la Recherche, as part of the
programme Investissements d'avenir under the reference ANR-11-IDEX-0004-02, and the Labex ``Institut Lagrange de Paris'' (ANR-11IDEX-0004-02,  ANR-10-LABX-63) which in particular funded the scholarship of SLW. MDG acknowledges the support of the Agence Nationale de Recherche grant ANR-15-CE31-0002 ``HiggsAutomator.'' HRG is funded by the Consejo Nacional de Ciencia y Tecnolog\'{i}a, CONACyT, scholarship no.\ 291169.

\bibliographystyle{JHEP}
\bibliography{references}

\end{document}